\pdfoutput=1
\documentclass[10pt,conference]{IEEEtran}
\IEEEoverridecommandlockouts
\usepackage[dvipdfmx]{graphicx}
\usepackage{caption}
\usepackage{latexsym}
\usepackage{physics}
\usepackage{bm}
\usepackage{algorithmic}
\usepackage{algorithm}
\usepackage{xcolor}
\usepackage{amsmath,amssymb,amsfonts}
\usepackage[style=ieee]{biblatex}
\usepackage{textcomp}
\usepackage{url}
\usepackage{tikz}
\usetikzlibrary{quantikz2}
\def\BibTeX{{\rm B\kern-.05em{\sc i\kern-.025em b}\kern-.08em
    T\kern-.1667em\lower.7ex\hbox{E}\kern-.125emX}}
\begin{document}

\renewcommand{\algorithmicrequire}{\textbf{Input:}}
\renewcommand{\algorithmicensure}{\textbf{Output:}}
\newcommand{\comment}[1]{\textcolor{red}{#1}}

\title{Efficient Internal Strategies in Quantum Relaxation based Branch-and-Bound\\
 \thanks{This work was performed for Council for Science, Technology and Innovation (CSTI), Cross-ministerial Strategic Innovation Promotion Program (SIP), “Promoting the application of advanced quantum technology platforms to social issues”(Funding agency : QST).}}

\author{\IEEEauthorblockN{Hiromichi Matsuyama}
\IEEEauthorblockA{\textit{Jij Inc.} \\
Tokyo, Japan \\
Email: h.matsuyama@j-ij.com}
\and
\IEEEauthorblockN{Wei-hao Huang}
\IEEEauthorblockA{\textit{Jij Inc.} \\
Tokyo, Japan \\
Email: w.huang@j-ij.com}
\and
\IEEEauthorblockN{Kohji Nishimura}
\IEEEauthorblockA{\textit{Jij Inc.} \\
Tokyo, Japan \\
Email: knishimura@j-ij.com}
\and
\IEEEauthorblockN{Yu Yamashiro}
\IEEEauthorblockA{\textit{Jij Inc.} \\
Tokyo, Japan \\
Email: y.yamashiro@j-ij.com}
}

\maketitle

\begin{abstract}
A combinatorial optimization problem is to find an optimal solution under the constraints. 
This is one of the potential applications for quantum computers.
Quantum Random Access Optimization (QRAO) is the quantum optimization algorithm that encodes multiple classical variables into a single qubit to construct a quantum Hamiltonian, thereby reducing the number of qubits required.
The ground energy of the QRAO Hamiltonian provides a lower bound on the original problem's optimal value before encoding. 
This property allows the QRAO Hamiltonian to be used as a relaxation of the original problem, and it is thus referred to as a quantum relaxed Hamiltonian.
In the Branch-and-Bound method, solving the relaxation problem plays a significant role.
In this study, we developed Quantum Relaxation based Branch-and-Bound (QR-BnB), a method incorporating quantum relaxation into the Branch-and-Bound framework. 
We solved the MaxCut Problem and the Travelling Salesman Problem in our experiments. 
In all instances in this study, we obtained the optimal solution whenever we successfully computed the exact lower bound through quantum relaxation.
Internal strategies, such as relaxation methods and variable selection, influence the convergence of the Branch-and-Bound. 
Thus, we have further developed the internal strategies for QR-BnB and examined how these strategies influence its convergence.
We show that our variable selection strategy via the expectation value of the Pauli operators gives better convergence than the naive random choice.
QRAO deals with only unconstrained optimization problems, but QR-BnB can handle constraints more flexibly because of the Branch-and-Bound processes on the classical computing part.
We demonstrate that in our experiments with the Travelling Salesman Problem, the convergence of QR-BnB became more than three times faster by using the information in the constraints.
\end{abstract}


\section{Introduction}
Combinatorial optimization problems involve finding a solution that minimizes (or maximizes) an objective function under constraints~\cite{korte2011combinatorial, Laurence_Wolsey2020-gj, conforti2014integer}. 
Various problems, such as the Vehicle Routing Problem, Crew Scheduling Problem, and Facility Location Problem, can be formulated as Combinatorial optimization problems. 
Therefore, various classical algorithms have been proposed to obtain an optimal or practically useful solution.
The branch-and-bound method is a classical exact solution algorithm~\cite{Land1960-ct, Morrison2016-dw}. 
Its techniques are widely used in many commercial and non-commercial software programs, such as Gurobi~\cite{gurobi} and CBC~\cite{john_forrest_2023_10041724}.

Recent developments in both hardware and software in quantum computation have led to various potential applications.
One application that draws attention is the combinatorial optimization problem~\cite{Abbas2023-aj}. 
Current quantum devices, which lack error correction and have fewer than a few hundred physical qubits, are called Noisy Intermediate Scale Quantum (NISQ) devices~\cite{Preskill2018-jq, Bharti2022-dx}. 
The Quantum Approximate Optimization Algorithm (QAOA)~\cite{Farhi2014-lj, Blekos2023-er} is a NISQ algorithm for combinatorial optimization inspired by adiabatic quantum computation~\cite{kadowaki1998, Farhi2000-fs, Albash2018-nt}.
QAOA encodes a combinatorial optimization problem into an Ising Hamiltonian and searches its ground state as an optimal solution.
QAOA and its variants have been applied to various NP-hard problems~\cite{Blekos2023-er} such as the MaxCut problem~\cite{Farhi2014-lj, Chalupnik2022qce} and the graph coloring problem~\cite{tabi2020qce, Bravyi2022hybridquantum}.

QAOA has some issues as a method for solving combinatorial optimization problems.
The first issue is the number of qubits required to describe the problem. 
We need as many qubits as the number of classical bits in the mathematical model.
Several quantum optimization algorithms have been proposed to reduce the number of required qubits~\cite{Fuller2021-su, Tan2021-fh, Rancic2023-ju, sciorilli2024largescale}. 
Quantum Random Access Optimization (QRAO) is a method to reduce the number of qubits by encoding multiple classical bits into a small number of qubits using a Quantum Random Access Code (QRAC)~\cite{Fuller2021-su, Teramoto2023-ix, Teramoto2023-jd}. 
QAOA only uses the Pauli $Z$ operators to encode the problem into the Hamiltonian. 
However, in QRAO, Pauli $X, Y$ operators are also used to encode.
Due to the difference between the QRAO Hamiltonian $\tilde{H}$ and the Ising Hamiltonian $H$, these ground states are different. 
The ground state of $\tilde{H}$ can be an entangled quantum state.
Therefore, a rounding algorithm must be applied to obtain a classical solution from the ground state of $\tilde{H}$. 
Fuller \textit{et al.} showed that QRAO provides an approximate solution with the approximation ratio under an appropriate Rounding algorithm~\cite{Fuller2021-su}.
Moreover, they showed that the ground energy of $\tilde{H}$ becomes a lower bound on the original problem.

The second issue with QAOA is the handling of constraints. 
The constrained optimization problem must be converted to Quadratic Unconstrained Binary Optimization (QUBO) using the relaxation method, such as the Penalty method, to encode into the Ising Hamiltonian.
Relaxation of constraint is a problem-independent technique, but it requires tuning weights of penalty terms.
Since tuning weights is an optimization problem, tuning weights in problems with many constraints becomes challenging.
Quantum Alternating Operator Ansatz has been proposed to overcome this problem~\cite{Hadfield2019-lb}.
However, it requires the design of a problem-specific quantum circuit, which tends to increase circuit depth.

The ground energy of the QRAO Hamiltonian $\tilde{H}$ gives a lower bound on the original problem.
The problem is known as quantum relaxation.
Solving relaxation problems is a major subroutine of the branch-and-bound method, the classical exact solution method.
In~\cite{huang2023qce}, we proposed the concept of Quantum Relaxation based Branch-and-Bound (QR-BnB), a quantum-classical hybrid algorithm using quantum relaxation within the framework of classical branch-and-bound methods, and showed that QR-BnB works by conducting experiments on a single small instance of the MaxCut problem.
QR-BnB consists of two fundamental processes: the branching process, which divides the problem into subproblems, and the bounding process, which evaluates the subproblem by solving the quantum relaxation problem and prunes the subproblem tree.
The branching and bounding processes are performed on the classical device, and the quantum device is used only to solve the subproblem's quantum relaxation problem.
QR-BnB requires less qubits than QAOA due to quantum relaxation.

In this paper, we developed QR-BnB to handle combinatorial optimization problems more efficiently. 
It is known that the performance of the branch-and-bound algorithm depends on various internal strategies, such as what kind of relaxation method is applied, how to select fixing variables, and how to branch the subproblem tree.
In this study, we examined several internal strategies to investigate the appropriate strategy for QR-BnB through detailed experiments using the MaxCut, an unconstrained optimization problem, and the Travelling Salesman Problem, a constrained optimization problem.
The optimal solution was obtained for all problem instances when an exact lower bound on quantum relaxation was obtained.
In this case QR-BnB provides the optimal solution with guarantees.
We developed variable selection strategies using the expectation value of Pauli operators.
Our variable selection strategies always give better convergence than Random selection of the fixing variables.
We further examine the treatment of constraints in QR-BnB by branching strategies using constraints information.
In particular, QR-BnB converged faster when using a branching strategy with constraints information than a branching strategy without constraints information.

We also performed experiments where the lower bound of the relaxation solution was computed using the Variational Quantum Eigensolver (VQE)~\cite{Peruzzo2014-kq} to evaluate its availability for NISQ devices.
In this case, QR-BnB cannot guarantee optimality due to the heuristic nature of VQE.
However, we found that QR-BnB provides an optimal solution in almost all cases, even in the VQE case.

The structure of this paper is as follows: In Sec. \ref{sec:qrao}, we explain QRAO which is the essential component of this study. 
Sec. \ref{sec:bnb} explains QR-BnB and its development in this study.
We discuss the experiment setting at Sec. \ref{sec:setting} and the experimental results using QR-BnB for the Maxcut problem and TSP at Sec. \ref{sec:result}.
Finally, in Sec. \ref{sec:summary}, we summarize our results and discuss future works.

\section{Quantum Random Access Optimization\label{sec:qrao}}
Quantum Random Access Optimization (QRAO)~\cite{Fuller2021-su, Teramoto2023-ix, Teramoto2023-jd} is a quantum optimization algorithm that utilizes Quantum Random Access Codes (QRAC)~\cite{Ambainis1999-cg, Ambainis2002-sb}. 
The number of qubits can be reduced compared to the conventional quantum optimization algorithm using the Ising Hamiltonian~\cite{Fuller2021-su}. 
QRAO consists of two steps.
In the first step, we construct a quantum relaxed Hamiltonian $\tilde{H}$ using QRAC and search for its ground state.
In the second step, we apply the rounding algorithm to the quantum state $\tilde{\rho}$ to obtain the classical solution.
QRAO has been proposed for the (weighted) MaxCut problem~\cite{Fuller2021-su}. 
However, its construction can be applied to the general Ising Hamiltonian, so we use the general Ising Hamiltonian for the explanation.

\subsection{Quantum Relaxed Hamiltonian \label{subsec:relaxed_hamiltonian}}
This subsection explains the construction of the relaxed Hamiltonian using $(3,1,p)$-QRAO as an example. 
$(3,1,p)$-QRAC encodes three classical bits $x_1, x_2, x_3 \in \{0,1\}$ into a single qubit as
\begin{equation}
\begin{split}
    \rho&(x_1,x_2,x_3) \\
    &= \frac{1}{2}\left(I + \frac{1}{\sqrt{3}}\left((-1)^{x_1}X + (-1)^{x_2}Y + (-1)^{x_3}Z\right)\right).
\end{split}
\label{eq:(3,1)-qrac}
\end{equation}
Three classical bits are encoded by mapping them to the Pauli $X, Y, Z$  operators of a single qubit.
The success probability of decoding for $(3,1,p)$-QRAC is $p=\frac{1}{2}\left(1 + \frac{1}{\sqrt{3}}\right)\approx 0.79$.
At most $4^n-1$ classical bits can be encoded in $n$ qubits using QRAC with $p \geq 1/2$, so $(3,1,p)$-QRAC is the maximum encoding for a single qubit~\cite{Hayashi2006-rr}.

Next, we describe the construction of the relaxed Hamiltonian. 
In QAOA, optimization problems are encoded into the Ising Hamiltonian as below:
\begin{equation}
H = \sum_{ij} J_{ij}Z_iZ_j + \sum_i h_i Z_i.
\label{eq:ising_hamiltonian}
\end{equation}
Since encoding uses only the Pauli $Z_i$ operators, $H$ is diagonal, and its ground state corresponds to the classical optimal solution.
In QRAO, not only Pauli $Z_i$ operators but also Pauli $X_i, Y_i$ operators are used to construct a relaxed Hamiltonian $\tilde{H}$.
Since QRAO encodes multiple classical bits into a single qubit to construct $\tilde{H}$, we must first determine which classical bits to map to which qubit.
We define the interaction graph $G(E, V)$ as an undirected graph composed of nodes $v_i,v_j\in V$ connected by edges such that $J_{ij}\neq 0$.
The QRAO mapping requires that adjacent classical bits on $G(E, V)$ be encoded in different qubits.
We can determine the mappings by solving the vertex coloring problem on $G(E, V)$.
We can employ the greedy method to solve the vertex coloring problem since obtaining a feasible solution rather than an optimal one is sufficient.
By solving the coloring problem, nodes $v_i,v_j$ at both ends of edge $(i,j)\in E$ have different colors $c(i)\neq c(j)$.
Then, we assign up to three nodes $v_i\in V_c$ with each color $c\in C$ to each qubit.
Consequently, the quantum relaxed Hamiltonian $\tilde{H}$ corresponding to eq.~\eqref{eq:ising_hamiltonian} becomes
\begin{equation}
\tilde{H} = \sum_{i,j} 3J_{ij}P_{i}P_{j} + \sum_i \sqrt{3}h_i P_i,
\label{eq:relax_hamiltonian}
\end{equation}
where $P_i \in \{X_i, Y_i, Z_i\}$ is the Pauli operator corresponding to node $v_i$. 
The coefficient $3, \sqrt{3}$ appears for $\tilde{\rho}_\mathrm{opt}$, the state in which the optimal solution is encoded using $(3,1,p)$-QRAC, to satisfy $\mathrm{Tr}(\tilde{H}\tilde{\rho}_\mathrm{opt})= OPT$ where $OPT$ is the optimal solution to the original problem.
Since $\tilde{H}$ is not diagonal, the ground state $\tilde{\rho}$ can be an entangled quantum state.
Thus, $\tilde{\rho}$ does not coincide with the optimal solution of the original problem. 
There is also no guarantee that $\tilde{\rho}$ becomes $\tilde{\rho}_\mathrm{opt}$.
Interestingly, however, $\mathrm{Tr}(\tilde{H}\tilde{\rho})$ gives a lower bound of the optimal value on the original problem.

The number of qubits required for each color is $n_c = \lceil |V_c|/3\rceil$, which means that up to three times more compression ratio can be achieved than in eq.~\eqref{eq:ising_hamiltonian}.
Note that the compression ratio depends on the given Ising Hamiltonian.
In particular, if the given Ising Hamiltonian is fully connected, no qubits are reduced.

The relaxed Hamiltonian for $(2,1,p)$-QRAO can be constructed similarly. 
The only difference is the number of qubits to be encoded in a single qubit and the coefficient of each term, and the relaxed Hamiltonian for $(2,1,p)$-QRAO is
\begin{equation}
\tilde{H} = \sum_{i,j} 2J_{ij}P_{i}P_{j} + \sum_i \sqrt{2}h_i P_i,
\label{eq:21relax_hamiltonian}
\end{equation}
where $P_i \in \{X_i, Z_i\}$.
Recently, other QRAOs have also been proposed~\cite{Teramoto2023-ix}.

\subsection{Rounding Algorithms}
Since the ground state $\tilde{\rho}$ of $\tilde{H}$ can be an entangled quantum state, we need rounding algorithms to decode a classical solution from $\tilde{\rho}$. 
Fuller \textit{et al.} proposed two rounding algorithms, Magic State Rounding and Pauli Rounding~\cite{Fuller2021-su}. 
The Magic State Rounding is a method that decodes information from three classical bits at once. 
This method has been proven to have an approximation ratio of $5/9$ for $(3,1,p)$-QRAC and $5/8$ for $(2,1,p)$-QRAC in the MaxCut problem.

Pauli Rounding is a method that decodes classical information by measuring the expected value $\mathrm{Tr}[P_i \tilde{\rho}]$ of the Pauli operator $P_i$ used to encode each classical variable for all classical variables $i \in V$.
Then, the sign is taken as the value corresponding to the variable.
Pauli Rounding has been reported to yield better solutions in most cases compared to Magic State Rounding by the numerical experiments~\cite{Teramoto2023-jd}. 
However, unlike Magic State Rounding, Pauli Rounding does not have approximation guarantees, even in the case of the MaxCut problem.

\section{Quantum Relaxation based Branch-and-Bound Algorithm\label{sec:bnb}}
In this section, we describe the Quantum Relaxation based Branch-and-Bound (QR-BnB) algotirhm and the strategy of the algorithms examined in this study. 
The branch-and-bound method is an algorithm for obtaining optimal solutions to the combinatorial optimization problems~\cite{Land1960-ct, Laurence_Wolsey2020-gj, Morrison2016-dw}.
Solving the relaxation problem is a major subroutine of the branch-and-bound method.
QR-BnB uses quantum relaxation as its relaxation problem.
In QR-BnB, quantum devices are used only to solve relaxation problems, while classical devices are used for constraint-dependent processing and solution construction.
\begin{figure}[!t]
\begin{algorithm}[H]
    \caption{Quantum Relaxation based Branch-and-Bound with strategies}
    \label{alg1}
    \begin{algorithmic}[1]    
    \REQUIRE Original Problem $S$
    \ENSURE optimal solution $\bm{x}_\mathrm{inc}$
    \STATE $L \leftarrow \{S\}, z_\mathrm{inc} \leftarrow \infty$
    \WHILE{ $L \neq \varnothing$ }
        \STATE Take $S_\mathrm{sub}$ from $L$ using tree search strategy\label{top_of_while_loop}

        \STATE Encode $S_\mathrm{sub}$ into Relaxed Hamiltonian $\tilde{H}$
        \STATE $\tilde{z}^\mathrm{sub}_\mathrm{QR} \leftarrow$ ground energy of $\tilde{H}$
        \STATE $z,\bm{x} \leftarrow$  objective value and state by Pauli Rounding

        \IF{$z < z_\mathrm{fes}$ and $\bm{x}$ is feasible}
            \STATE $z_\mathrm{inc} \leftarrow z,\ \bm{x}_\mathrm{inc} \leftarrow \bm{x}$
        \ENDIF 

        \IF{$z_\mathrm{fes} < \tilde{z}^\mathrm{sub}_\mathrm{QR}$}
            \STATE Go to line 2
        \ENDIF

        \STATE Choose index $i$ by using variable selection strategy
        \STATE Branch $S_\mathrm{sub}$ into set of subproblems using branching strategy and add to $L$
    \ENDWHILE
    \RETURN $\bm{x}_\mathrm{inc}$
    \end{algorithmic}
\end{algorithm}
\end{figure}

\subsection{Algorithm Flow\label{subsec:alg_flow}}
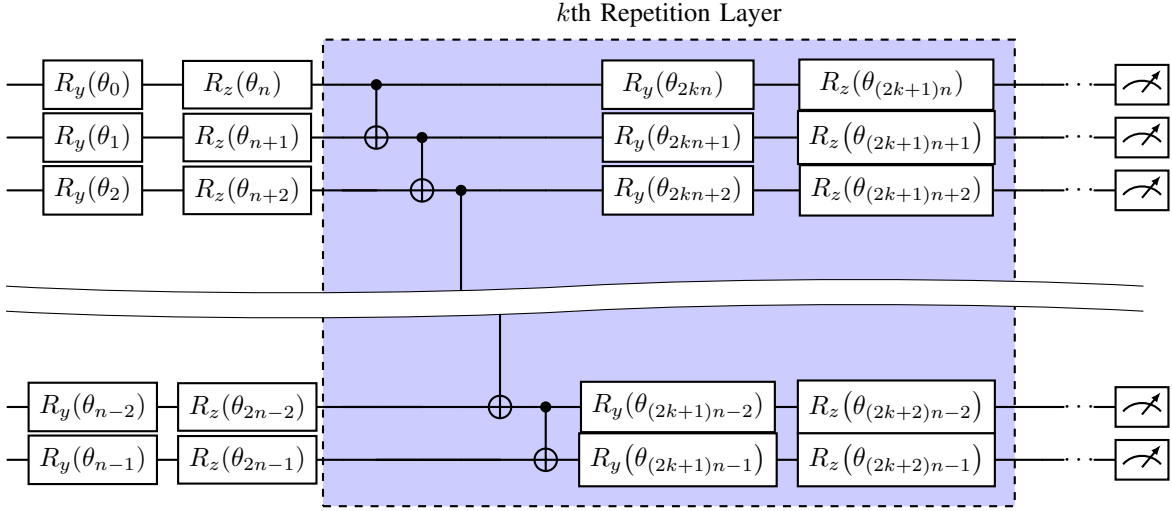
\begin{figure*}[!t]
\centering
\begin{quantikz}[column sep=8pt, row sep={20pt,between origins}]
    & \gate{R_y(\theta_0)} & \gate{\ R_z(\theta_n)\ \ } &[2pt]\gategroup[8,steps=8,style={dashed,fill=blue!20},background,label style={anchor=north,yshift=0.4cm}]{{
$k$th Repetition Layer}} &\ctrl{1}  &&&& & \gate{\ R_y(\theta_{2kn})\ \ } & \gate{\ R_z(\theta_{(2k+1)n})\ \ } &&[2pt]&\cdots&\meter{}\\
    & \gate{R_y(\theta_1)} & \gate{R_z(\theta_{n+1})} &&\targ{} &\ctrl{1}&&&&  \gate{R_y(\theta_{2kn+1})} & \gate{R_z\qty(\theta_{(2k+1)n+1})}&&&\cdots&\meter{}\\
    & \gate{R_y(\theta_2)} & \gate{R_z(\theta_{n+2})} &&\qw  & \targ{} &\ctrl{2} &&  & \gate{R_y(\theta_{2kn+2})} & \gate{R_z(\theta_{(2k+1)n+2})}&&&\cdots&\meter{}\\
    [1pt]\\
    \wave&&&&&&&\ctrl{2}&&&&&&&&&&&&&\\
    [1pt]\\
    & \gate{R_y(\theta_{n-2})} & \gate{R_z(\theta_{2n-2})} &&\qw  &\qw && 
    \targ{} &\ctrl{1} & \gate{R_y(\theta_{(2k+1)n-2})} & \gate{R_z\qty(\theta_{(2k+2)n-2})}&&&\cdots&\meter{}\\
    & \gate{R_y(\theta_{n-1})} & \gate{R_z(\theta_{2n-1})} &&&\qw  &\qw & \qw & \targ{}  & \gate{R_y\qty(\theta_{(2k+1)n-1})} & \gate{R_z\qty(\theta_{(2k+2)n-1})}&&&\cdots&\meter{}
\end{quantikz}
\caption{Variational quantum circuit of the Hardware-Efficient Ansatz used in the experiments.}
\label{fig:ansatz}
\end{figure*}

QR-BnB employs quantum relaxation within the branch-and-bound method. 
The pseudocode for QR-BnB is shown in Fig. \ref{alg1}. 
The algorithm consists of a branch process that generates a subproblem tree and a bound process that prunes the subproblem tree.
We consider an optimization problem with $n$ binary variables and $m$ constraints:
\begin{equation}
S:\ z^* = \min \{\bm{x}^TQ\bm{x}:\bm{x} \in X\},
\label{eq:quadratic_binary_problem}
\end{equation}
where $X = \{\bm{x} \in \{0,1\}^{n} : A\bm{x} \leq \bm{b}\}$ is the feasible set of the problem, $\bm{b} \in \mathbb{R}^m, A\in \mathbb{R}^{m\times n}, Q\in \mathbb{R}^{n\times n}$.

First, we explain the branching process. 
The branching process divides the feasible set into smaller subsets that do not overlap.
In particular, if the problem has only binary variables, this process generates two subproblems by fixing a variable:
\begin{equation}
    \begin{split}
        S_0:\ &z_0^* = \min\{\bm{x}^TQ\bm{x}:\bm{x} \in X_0\},\\
        S_1:\ &z_1^* = \min\{\bm{x}^TQ\bm{x}:\bm{x} \in X_1\},\\
    \end{split}
\end{equation}
where $X_0 = X \cap \{\bm{x}: x_i = 0\}, X_1 = X \cap \{\bm{x}: x_i = 1\}$. 
We can repeat this process to generate smaller subproblems by fixing additional variables from these $X_0, X_1$.
The relation between subproblems can be drawn as a tree.
The root node of this subproblem tree is the original problem $S$, and the leaves correspond to the solutions to the problem.
During the search process, the solution to the problem $S$ is found, and it is stored as an incumbent solution $\bm{x}_\mathrm{inc}$ for the algorithm.
The final incumbent solution is the output of the algorithm.

The branching process expands the subproblem tree, which is no different from a brute-force search.
The bound process reduces the size of the subproblem tree by evaluating and pruning the tree.
The subproblem $S_\mathrm{sub}$ is evaluated by the optimal value of quantum relaxation $\tilde{z}^\mathrm{sub}_\mathrm{QR}$. 
We denote the objective value of the feasible and optimal solution of the original problem $S$ as $z_\mathrm{fes}$, $z^*$ respectively, and the optimal value of the relaxation problem as $\tilde{z}_\mathrm{QR}$.
As mentioned in section \ref{subsec:relaxed_hamiltonian}, $\tilde{z}_\mathrm{QR}$ gives a lower bound on the optimal value of $S$.
$z_\mathrm{fes}$ gives an upper bound on the optimal value of $S$. 
Therefore, relation $\tilde{z}_\mathrm{QR} \leq z^* \leq z_\mathrm{fes}$ holds.
On the other hand, since the feasible set of the subproblem is more bounded than $S$, the objective value of the relaxed subproblem $\tilde{z}^\mathrm{sub}_\mathrm{QR}$ does not necessarily satisfy $\tilde{z}^\mathrm{sub}_\mathrm{QR} \leq z_\mathrm{inc}$ where $z_\mathrm{inc}$ is the objective value of the incumbent solution.
This implies that no solution in the feasible set of the relaxation problem for this subproblem is better than the incumbent solution.
Thus, we prune the branch below this subproblem.

One advantage of using quantum relaxation is that it can solve problems involving second-order objective functions, such as the eq.~\eqref{eq:quadratic_binary_problem}, which cannot be solved by simple linear relaxation.
Moreover, although the subproblems are evaluated by quantum relaxation, the search for the solution itself is performed using the branching process, so the difficulty of handling constraints in the quantum algorithm can be indirectly avoided by pruning infeasible solutions from the subproblem tree when searching for solutions.
A detailed discussion about the feasibility evaluations of the subproblem is in the Appendix.

\subsection{Strategies in Branch-and-Bound\label{subsec:strategies}}
In branch-and-bound methods, convergence can be very slow if we cannot prune the subproblem tree efficiently.
It is known that convergence speed depends on internal operations.
For this purpose, various methods have been proposed for internal operation strategies~\cite{Morrison2016-dw}.
In this study, we examined several strategies to determine how changes in these strategies affect QR-BnB's convergence.
We investigate three strategies: the variable selection strategy, the branching strategy, and the tree search strategy.
In this subsection, we describe these strategies and improvements specific to QR-BnB.

First, we describe the variable selection strategy for choosing the variables to be fixed.
In the case of linear relaxation, the solution is obtained as continuous values.
There are two simple variable selection methods that utilize continuous values.
In the Least Fractional rule, the variable closest to $1$ or $0$ is fixed.
On the other hand, the Most Fractional rule fixes the variable farthest from $1$ or $0$.
Unlike these methods, the Random rule selects variables to be fixed regardless of continuous values.
In the case of quantum relaxation, since the relaxed solution is in a quantum state, only the Random rule can be applied.
To overcome this issue, in this study, we developed a variable selection method through the expectation value $\mathrm{Tr}[P_i \tilde{\rho}]$ of the Pauli operator $P_i$ in the obtained quantum state $\tilde{\rho}$.
We use $\mathrm{Tr}[P_i \tilde{\rho}]$ as the continuous value in the linear relaxation
\footnote{$\mathrm{Tr}[P_i \tilde{\rho}]$ satisfies $-1\leq \mathrm{Tr}[P_i \tilde{\rho}] \leq 1$. 
The value closest to $0$ is fixed for the Most Fractional rule, and the values closest to $1$ and $-1$ are fixed for the Least Fractional rule.}.  
These expectation values allow QR-BnB to use the Least Fractional and Most Fractional rules as variable selection strategies.

Obtaining a better incumbent solution with $z_\mathrm{inc}$ close to the optimal value at an early stage of the algorithm is crucial for reducing the branches to be explored. 
In the naive application of QR-BnB, the solution to the original problem can only be obtained when we arrive at the leaves of the subproblem tree.
As we mentioned above, we computed the expectation value $\mathrm{Tr}[P_i \tilde{\rho}]$ for variable selection strategies.
Since taking the sign of $\mathrm{Tr}[P_i \tilde{\rho}]$ is Pauli Rounding, we apply Pauli Rounding to each subproblem of QR-BnB to obtain one solution for the original problem to increase the frequency of $z_\mathrm{inc}$ updates.

The branching strategy is a strategy for branching the subproblem tree.
The most basic branching strategy is the Binary Branch, which generates two subproblems by fixing one variable, as discussed in section \ref{subsec:alg_flow}.
However, for problems with specific constraints, we can branch the subproblem tree more efficiently with information about the constraints.
Let us consider the problem with the onehot constraint $\sum_{k\in K} x_k = 1$.
Once the variable $x_i$ in the onehot constraint is fixed to $1$, the other variables must be $0$.
Therefore, we can fix multiple variables simultaneously as $x_i=1,\ x_k=0\ (k\in K/\{i\})$.
In this study, we call this method the Onehot Branch.

The tree search strategy is a method for determining the search order of a subproblem tree.
A general tree search algorithm can be used for this.
In this study, we examined three search methods: Depth First Search (DFS), Breath First Search (BrFS), and Best First Search (BFS).

\begin{figure*}[!t]
    \centering
    \captionsetup{width=\linewidth}
    \includegraphics[width=\linewidth]{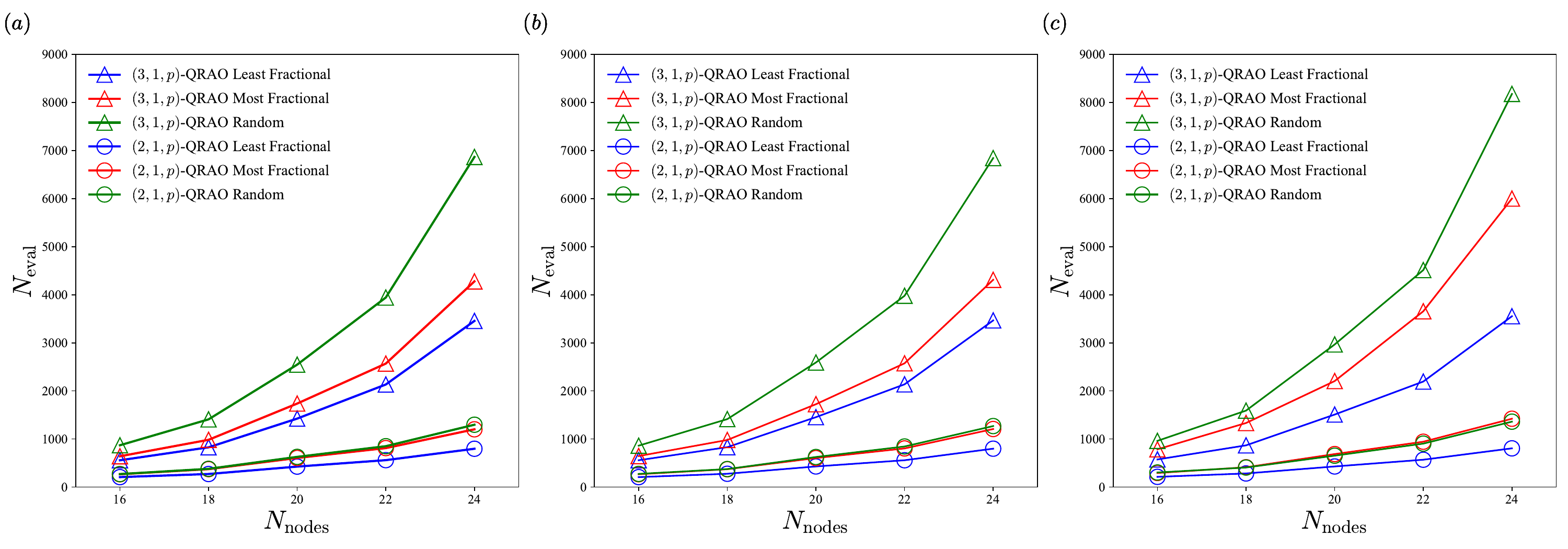}
    \caption{
    $N_\mathrm{nodes}$ dependence of $N_\mathrm{eval}$ for the Maxcut problem with QR-BnB-Exact. 
    (a) Best First Search. (b) Breath First Search. (c) Depth First Search, showing changes in convergence with different tree search strategies. 
    Each line color corresponds to the variable selection strategies, and each mark corresponds to the quantum relaxation methods. 
    $(2,1,p)$-QRAO with the Least Fractional rule converges fastest regardless of the search strategy chosen.}
    \label{fig:maxcut_exact_eigen}
\end{figure*}

\begin{figure*}[!t]
    \centering
    \captionsetup{width=\linewidth}
    \includegraphics[width=\linewidth]{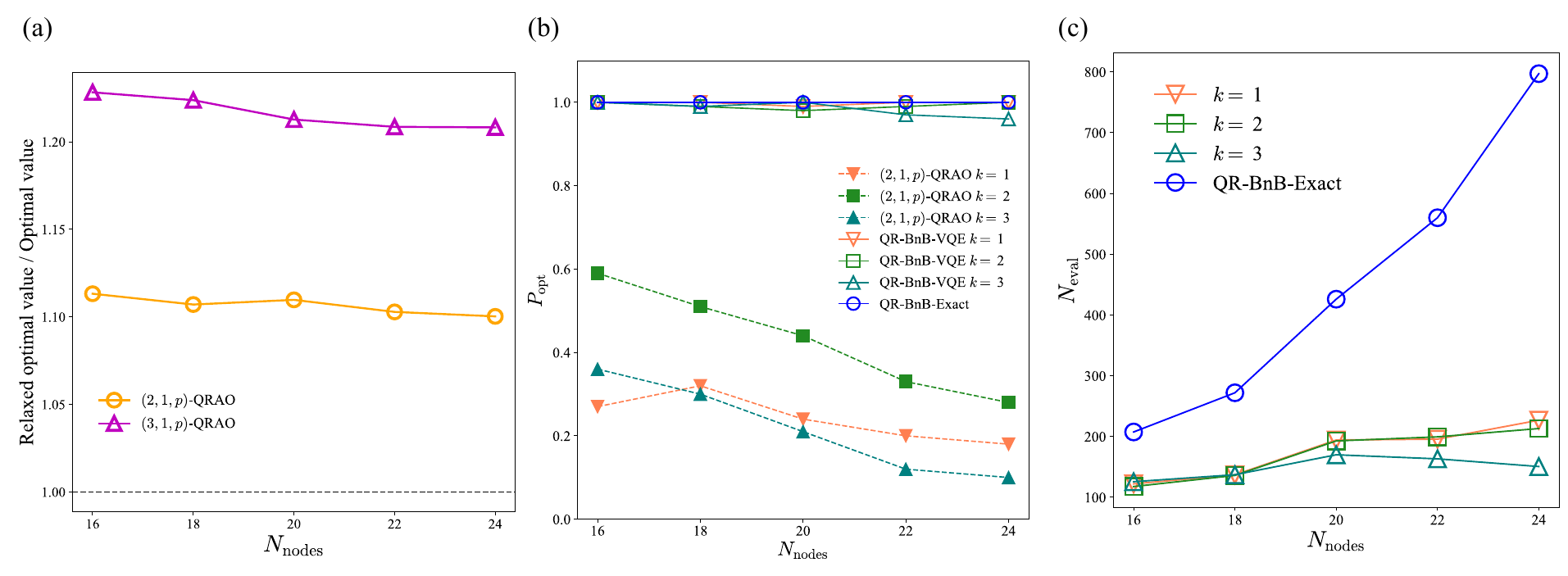}
    \caption{
    (a) Comparison of the size of the quantumness gap by quantum relaxation methods. 
    The quantumness gap of $(2,1,p)$-QRAO is always smaller than that of $(3,1,p)$-QRAO.
    (b) $N_\mathrm{nodes}$ dependences of $P_\mathrm{opt}$ for MaxCut problem with QR-BnB-VQE and naive $(2,1,p)$-QRAO.
    QR-BnB-VQE always gives better success probability than the naive QRAO.
    (c) $N_\mathrm{nodes}$ dependences of $N_\mathrm{eval}$ for MaxCut problem with QR-BnB-VQE.
    QR-BnB-VQE always provides better convergence than QR-BnB-Exact.}
    \label{fig:maxcut_gap_vqe}
\end{figure*}
\section{Experiment Settings \label{sec:setting}}
In this study, we solve the MaxCut problem and Travelling Salesman Problems (TSP) as examples of unconstrained and constrained optimization problems.
First, we describe these problems and their setup below.

MaxCut problem divides the node set $V$ of an undirected graph $G(V, E)$ into two subsets.
The problem is to maximize the sum of the number of edges where the nodes at both ends of the edges are assigned to different subsets.
The MaxCut problem on an undirected graph $G(V, E)$ is formulated by the following mathematical model:
\begin{equation}
    \max_{s}\quad \frac{1}{2}\sum_{(i,j)\in E} (1 - s_is_j),
\end{equation}
with the spin variable $s_i\in \{-1,1\}$ indicating the assignment to each subset.
Since this problem is a maximization problem, we reversed the sign of the problem and solved it as a minimization problem.
In this study, we evaluated the performance of QR-BnB by randomly generating $3$-regular graphs with $100$ samples, each from $16$ to $24$ nodes.

TSP is the problem in finding the shortest tour of $N$ cities.
This problem can be formulated as
\begin{equation}
\begin{split}
    \min_x\ &\sum_t \sum_{ij}d_{ij} x_{i,t} x_{j,t+1\ \mathrm{mod}\ N}\\
    \mathrm{s.t.}\ &\sum_{i=0}^{N-1} x_{i,t} = 1\quad \forall t \in \{0,1,\dots, N-1\},\\
    &\sum_{t=0}^{N-1} x_{i,t} = 1\quad \forall i \in \{0,1,\dots, N-1\},
\label{eq:tsp}
\end{split}
\end{equation}
using the variable $x_{i,t}\in \{0,1\}$, which is 1 when a salesman in a city $i$ at time $t$.
The problem has two onehot constraints for time and city, so that Onehot Branch can be applied.
QRAO requires transforming constrained optimization problems into the Ising Hamiltonians. 
Therefore, we transformed eq.~\eqref{eq:tsp} into the following QUBO:
\begin{equation}
\begin{split}
    \min\ &\sum_t \sum_{ij}d_{ij} x_{i,t} x_{j,t+1\ \mathrm{mod}\ N} \\
    &+ A\sum_t \left(\sum_{i=0}^{N-1} x_{i,t} - 1\right)^2 + B \sum_i \left(\sum_{t=0}^{N-1} x_{i,t} - 1\right)^2,
\end{split}
\end{equation}
using the Penalty method.
We then generated the Ising Hamiltonian by transforming the binary variables into spin variables.
We set $A = B = 1$ in our experiment. 
We evaluated the performance of QR-BnB using $100$ instances of $4$-city problems randomly distributed in $[0,1]^2$ as problem instances.

In this study, we changed the variable selection strategy, the branching strategy, the tree search strategy, and the quantum relaxation method to evaluate the convergence speed of QR-BnB.
We used two methods to calculate the lower bound by quantum relaxation in QR-BnB.
One is the computation of optimal lower bounds by exact diagonalization, and the other is by Variational Quantum Eigensolver (VQE)~\cite{Peruzzo2014-kq}.
We refer to the QR-BnB employing exact lower bound calculations as QR-BnB-Exact and the version using VQE as QR-BnB-VQE. 
We used hardware-efficient ansatz~\cite{Kandala2017-ya} for the variational quantum circuit of VQE.
It consists of parameterized RY, RZ gates and $CNOT$ entangling layers.
In this experiment, we used a linear entangler for the entangling layer (Fig. \ref{fig:ansatz}) and varied the number of layers $k$ from $1$ to $3$.
For the optimization of the parameter in the variational quantum circuit, we used the Nakanishi-Fuji-Todo (NFT) algorithm~\cite{Nakanishi2020-aq} which is the gradient free-type optimizer designed for hardware-efficient ansatz.
When the unfixed variables in QR-BnB were less than three bits, we performed a brute-force search without encoding into the quantum state.
We used Qamomile\footnote{https://github.com/Jij-Inc/Qamomile}, a quantum optimization library, to generate the quantum relaxed Hamiltonian, and qiskit~\cite{Qiskit} and QURI Parts\footnote{https://github.com/QunaSys/quri-parts}, an open source library for quantum computation to implement the algorithm.
To obtain the optimal solution of the problems for comparison, we linearized the problem employing the Glover-Woolsey method~\cite{Glover1973-hs} and solved it with CBC~\cite{john_forrest_2023_10041724}.

\begin{figure*}[!t]
    \centering
    \includegraphics[width=\linewidth]{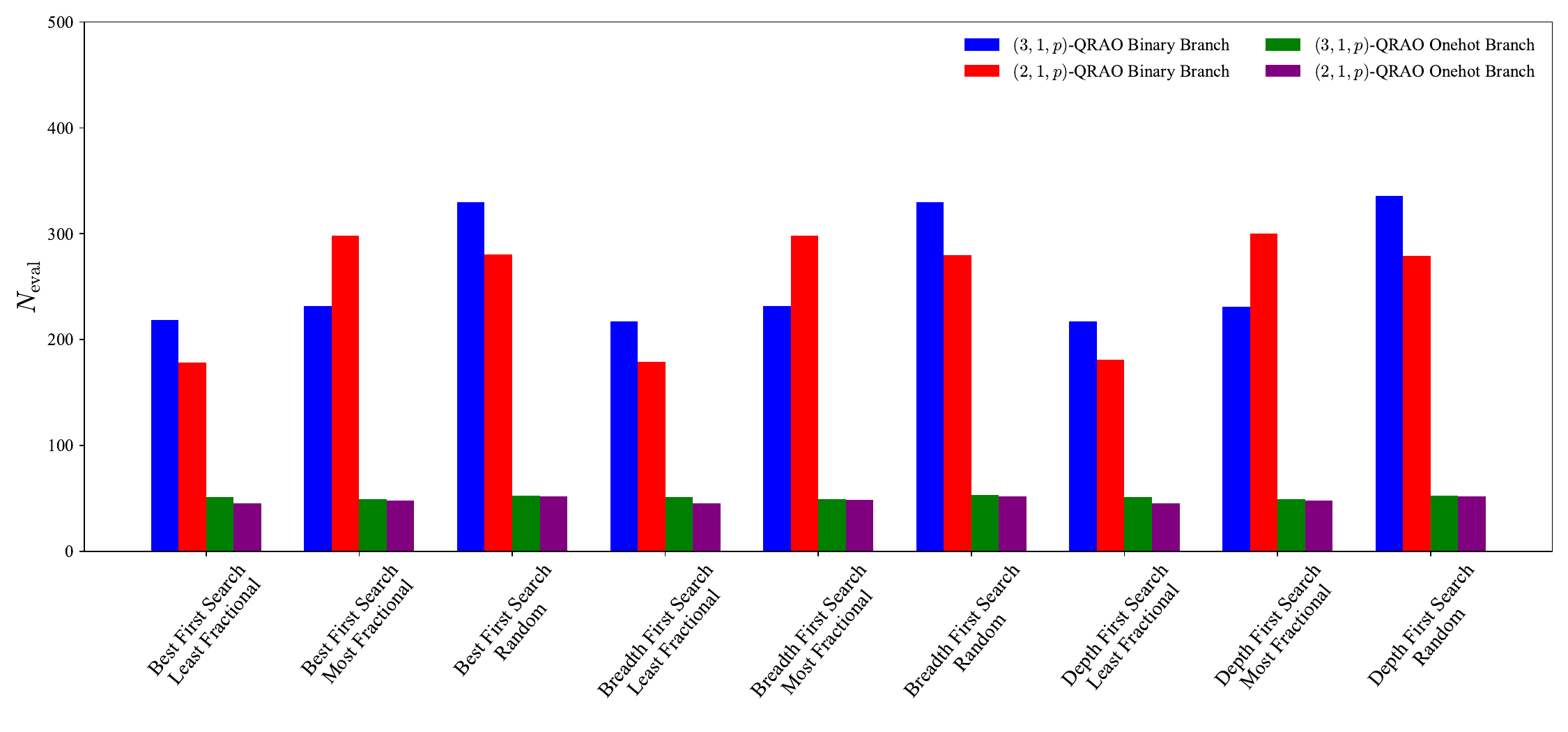}
    \caption{
    Comparison of $N_\mathrm{eval}$ for various strategies in TSP. 
    Using the Onehot Branch always provides faster convergence than the Binary Branch.}
    \label{fig:tsp_exact_eigen}
\end{figure*}
\section{Results \label{sec:result}}
\begin{figure*}[!t]
    \centering
    \includegraphics[width=\linewidth]{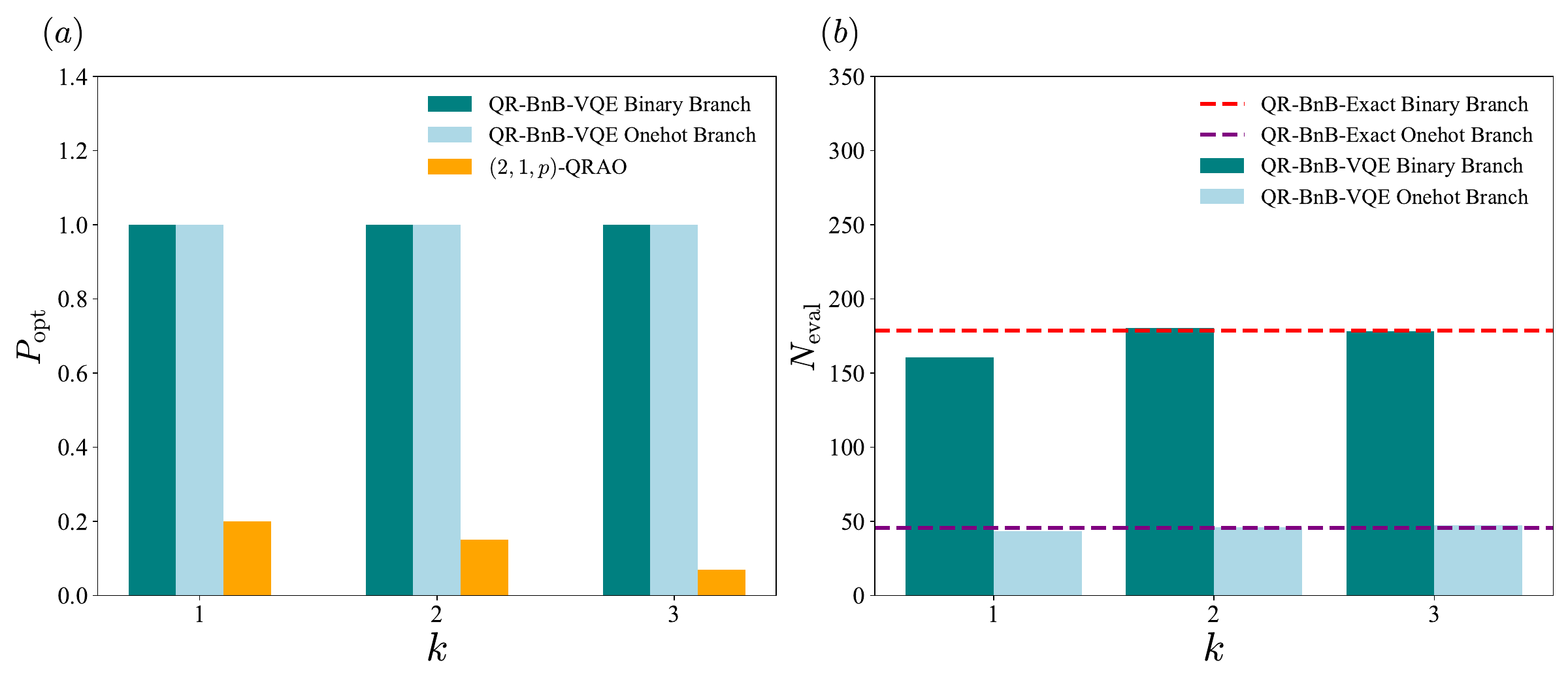}
    \caption{
    (a) $k$ dependeces for $P_\mathrm{opt}$ for TSP compared to QR-BnB-VQE and naive $(2,1,p)$-QRAO.
    QR-BnB-VQEs always provide an optimal solution.
    (b) $k$ dependeces for $N_\mathrm{eval}$ for TSP with QR-BnB-VQE.
    The Onehot Branch provides faster convergence than the Binary Branch.
    }
    \label{fig:tsp_vqe}
\end{figure*}

\subsection{MaxCut Problem\label{subsec:maxcut}}
First, we investigate the performance of QR-BnB on unconstrained optimization problems using the MaxCut problem.
We analyze the average number of evaluations of the quantum relaxation problems $N_\mathrm{eval}$ until converging the branch-and-bound process.
Figure \ref{fig:maxcut_exact_eigen} (a)-(c) show the dependence of $N_\mathrm{eval}$ on the problem size $N_\mathrm{nodes}$ when using QR-BnB-Exact.
QR-BnB-Exact gave optimal solutions for all problem instances for all methods.
Note that the optimality of the solutions is guaranteed by QR-BnB-Exact due to the branch-and-bound process.
(a), (b) and (c) are the results when BFS, BrFS, and DFS are employed as the tree search strategies, respectively. 
The color indicates the variable selection method, and the mark indicates the quantum relaxation method.
For $(2,1,p)$-QRAO, the dependence on the tree search strategy is small, but for $(3,1,p)$-QRAO, BFS converges faster than DFS.
Therefore, in the following, we will focus on the case of BFS (Fig. \ref{fig:maxcut_exact_eigen}(a)).
Next, we consider the dependence on the variable selection method. 
Regardless of the quantum relaxation method, the Least Fractional rule converges the fastest.
Similar results have been reported in the previous study of classical branch-and-bound algorithm~\cite{Ortega2003-io}.

Next, we focus on $N_\mathrm{eval}$ dependence on the quantum relaxation method.
$N_\mathrm{eval}$ for $(2,1,p)$-QRAO is always smaller than for $(3,1,p)$-QRAO, indicating that $(2,1,p)$-QRAO is always faster converging.
In the classical branch-and-bound method, the integrality gap, the ratio of the optimal values of the original problem and its linear relaxation problem, is related to convergence speed. 
In particular, convergence can be faster when the integrality gap is small.
Therefore, we consider the quantumness gap, which is the ratio of the optimal value of the original optimization problem and the quantum relaxation problem, to be similar in the case of linear relaxation.
As shown in Fig.~\ref{fig:maxcut_gap_vqe}~(a), the quantumness gap is larger for $(3,1,p)$-QRAO than for $(2,1,p)$-QRAO. 
As in the case of the integrality gap, the convergence speed varies with the size of the quantumness gap.
In the case of both quantum relaxation methods, the quantumness gap is independent of $N_\mathrm{nodes}$.

Next, we investigated the behavior of QR-BnB-VQE.
The optimal value of quantum relaxation $\tilde{z}_\mathrm{QR}$ and the expectation value obtained by VQE $\ev{\psi(\bm{\theta})|\tilde{H}|\psi(\bm{\theta})}$ have the relation $\ev{\psi(\bm{\theta})|\tilde{H}|\psi(\bm{\theta})}\geq \tilde{z}_\mathrm{QR}$.
Here, $\bm{\theta}$ denotes the parameters of the variational quantum circuit, and the equal sign is held when the optimal parameter $\bm{\theta}^*$ is obtained.
Therefore, if $\bm{\theta}^*$ could not be obtained, there is a possibility that the relaxed solution of the subproblem which should be $\tilde{z}_\mathrm{QR}\leq z_\mathrm{fes}$ is evaluated as $\ev{\psi(\bm{\theta})|\tilde{H}|\psi(\bm{\theta})}\geq z_\mathrm{fes}$.
This implies that the subproblems that should not be pruned are pruned.
Therefore, we evaluated the success probability $P_\mathrm{opt}$ of obtaining the optimal solution in QR-BnB-VQE in Fig.~\ref{fig:maxcut_gap_vqe} (b).
In this experiment, we used the fastest convergence setting in QR-BnB-Exact, the Least Fractional rule and BFS, and $(2,1,p)$-QRAO as quantum relaxation methods.
Because the convergence of QR-BnB-VQE may be slowed down when the optimal solution branches were pruned by mistake, we set the upper limit of the number of evaluations at $2000$ for $(2,1,p)$-QRAO terminated QR-BnB-VQE and evaluated the solutions obtained up to this point.
In all cases, QR-BnB-VQE has a higher success probability than naive $(2,1,p)$-QRAO.
$P_\mathrm{opt}$ for QRAO decreases with increasing $N_\mathrm{nodes}$, while that for QR-BnB-VQE is almost $N_\mathrm{nodes}$ independent.
In particular, for $k =1,2$, $P_\mathrm{opt}\geq 0.98$ in all problem sizes.
On the other hand, for $k=3$, $P_\mathrm{opt}$ tends to decrease as $N_\mathrm{nodes}$ increases.

Figure~\ref{fig:maxcut_gap_vqe} (c) shows the system size dependence of the number of evaluations required when an optimal solution was obtained using VQE.
The $N_\mathrm{eval}$ in QR-BnB-VQE is always less than in QR-BnB-Exact.
It is because $\ev{\psi(\bm{\theta})|\tilde{H}|\psi(\bm{\theta})}\geq \tilde{z}_\mathrm{QR}$, so the quantum relaxation gap is estimated to be small even when the lower bound is obtained.
Moreover, in QR-BnB-VQE, the dependence on the number of nodes is smaller than in QR-BnB-Exact.

\subsection{Travelling Salesman Problem}
\label{subsec:tsp}
We also solve TSP using QR-BnB as an example of a constrained optimization problem.
Since TSP has onehot constraints, we can apply Onehot Branch and Binary Branch.
We first investigated how these branching methods affect the performance of QR-BnB-Exact in Fig. \ref{fig:tsp_exact_eigen} .
As with the MaxCut problem, optimal solutions were obtained for all problem instances in this experiment.
Figure \ref{fig:tsp_exact_eigen} compares convergence between Binary Branch and Onehot Branch. 
The horizontal axis indicates the combination of tree search strategies and variable selection strategies. 
Each bar represents $N_\mathrm{eval}$ when using Binary Branch and Onehot Branch for $(3,1,p)$-QRAO and $(2,1,p)$-QRAO, respectively.
In all settings, convergence was improved when using Onehot Branch compared to Binary Branch. 
In particular, even in the case of BFS and Least Fractional rule, where Binary Branch was the fastest, using Onehot Branch was more than three times faster.
It is attributed to the fact that more variables can be fixed in a single branching operation, reducing the size of the branches that need to be explored.
In the Binary Branch, as in the MaxCut problem, $(2,1,p)$-QRAO requires fewer evaluations for convergence than $(3,1,p)$-QRAO in most cases.
However, interestingly, in the case of the Most Fractional rule, $(3,1,p)$-QRAO provides faster convergence.
Therefore, $N_\mathrm{eval}$ does not depend on the quantumness gap but rather on the interplay between the selected internal strategies and the gap.
On the other hand, in Onehot Branch, $N_\mathrm{eval}$ is almost the same, indicating that the influence of the branching method is more significant than that of the quantum relaxation method.

Finally, we present the results of solving the TSP using QR-BnB-VQE. 
Here, the experiment was conducted using the Least Fractional rule, BFS, and $(2,1,p)$-QRAO as a quantum relaxation method, which had the fastest convergence of the Binary Branch and Onehot Branch in QR-BnB-Exact.
Figure~\ref{fig:tsp_vqe} (a) shows the dependence of the success probability $P_\mathrm{opt}$ of QR-BnB-VQE and VQE for naive $(2,1,p)$-QRAO on the number of layers $k$.
In the case of QR-BnB-VQE, we obtain the optimal solution for both Onehot and binary branches in all instances.
However, in the case of naive $(2,1,p)$-QRAO, $P_\mathrm{opt}$ decreases with increasing $k$.
Next, we show $N_\mathrm{eval}$ for QR-BnB-VQE in Fig.~\ref{fig:tsp_vqe} (b).
As with QR-BnB-Exact, the Onehot Branch provides faster convergence than the Binary Branch in QR-BnB-VQE, indicating that the Onehot Branch improves convergence speed in both cases.
Moreover, $N_\mathrm{eval}$ for QR-BnB-VQE is similar to that of QR-BnB-Exact, suggesting that QR-BnB-Exact and QR-BnB-VQE exhibit similar behavior when solving TSP instances.

\section{Summary \label{sec:summary}}
QR-BnB is an algorithm that combines quantum relaxation and branch-and-bound methods.
QR-BnB provides a guaranteed optimal solution if we can obtain the exact lower bound of the subproblem through the quantum relaxation problem.
The quantum device only solves the quantum relaxation problem, so we only need fewer qubits compared to QAOA.
In this study, we developed QR-BnB by examining several internal strategies, particularly branching methods using constraints and variable selection methods through the expectation value of Pauli operators.
Moreover, we performed detailed experiments on these internal strategies to investigate the dependence of QR-BnB convergence on internal strategies.
In Sec. \ref{sec:result}, we solved the MaxCut problem and TSP using QR-BnB.
Using QR-BnB-Exact, which evaluates the exact lower bounds for each subproblem, optimal solutions were obtained in all the experiments in this study.
In our experiment, $(2,1,p)$-QRAO with a smaller quantumness gap converges faster than $(3,1,p)$-QRAO in most settings.
We have examined several variable selection strategies, and the Least Fractional rule we developed in this study provides better convergence than the Random rule.
In a comparison of branching strategies, the Onehot Branch, which uses constraint information, is more than three times faster than the Binary Branch in solving TSP.
In this study, we have also examined QR-BnB-VQE, which solves the quantum relaxation problem using VQE.
Due to the heuristic nature of VQE, QR-BnB-VQE cannot guarantee the optimality of the solution.
However, even in QR-BnB-VQE, we have obtained the optimal solutions in most instances.
Moreover, the convergence of QR-BnB-VQE is faster than that of QR-BnB-Exact.

We only considered relatively simple internal strategies of the classical side to evaluate the performance of QR-BnB.
On the other hand, for practical use, efficiency can be improved by incorporating more complicated strategies on the classical side of branch-and-bound methods and combining continuous and quantum relaxation.

One major direction for future study is to realize a method for systematically reducing the quantumness gap, similar to the cutting plane~\cite{dantzig1954solution, gomory1958outline}, which can reduce the integrality gap.
The classical solver is based on the Branch-and-Cut algorithm\cite{padberg1991branchandcut}, which combines the cutting plane method and Branch-and-Bound.
Therefore, a more efficient branch-and-cut solver can be established if a more efficient cutting plane method can be developed for quantum relaxation rather than linear relaxation.
Another major direction is to realize a more suitable quantum relaxation method.
In this study, we have used a Hamiltonian constructed of QRAOs as the quantum relaxation. 
It should be noted that a sufficient condition for quantum relaxation is to provide a lower bound on the original problem and that QRAO is not the only option for quantum relaxation.
Therefore, it is expected to develop a new quantum relaxation that can encode more qubits while keeping the quantumness gap small.

\section*{Acknowledgment}
H.M. thanks for the fruitful discussion with Kosei Teramoto and Rudy Raymond.

\appendix[Feasibility Evaluation]
\label{appendix:feasible}
In the branch-and-bound method, the constraints are evaluated at the branch process, and subproblems that do not satisfy the constraints can be pruned.
In this appendix, we discuss how to evaluate constraint violation.
We concentrate our discussion on linear equality constraints $\sum_iA_{ri}x_i = b_r$ or linear inequality constraints $\sum_iA_{ri}x_i \leq b_r$ based on eq.~\eqref{eq:quadratic_binary_problem}.

Let $I$ be the set of indices of all variables in the problem, and let $J\subset I$ be the set of indices of fixed variables. 
The linear constraint $\sum_{i\in I}A_{ri}x_i = b_r$ can then be rewritten as $\sum_{i\in I\setminus J}A_{ri}x_i = b_r - \sum_{j\in J}A_{rj}x_j=\tilde{b}_r$.
Here, $\tilde{b}_r$ is the constant part, including the constants created by fixing the variables.
Thus, the linear constraint retains its form for fixed variables.
Therefore, in the following, all constraints are assumed to be after the variables are fixed, and the sums are taken for $I\setminus J$.
We also define the upper and lower bounds of $\sum_iA_{ri}x_i$ for each constraint as $\sup(\sum_iA_{ri}x_i) = \sum_{i | A_{ri} > 0}A_{ri}$ and $\inf(\sum_iA_{ri}x_i) = \sum_{i | A_{ri} < 0}A_{ri}$ respectively.

First, consider the equality constraint $\sum_iA_{ri}x_i = \tilde{b}_r$.
$\tilde{b}_r$ must be between the upper and lower bounds that can be made with unfixed variables to satisfy the constraint condition.
This implies
\begin{equation}
\inf\qty(\sum_iA_{ri}x_i)  \leq \tilde{b}_r \leq \sup\qty(\sum_iA_{ri}x_i).
\end{equation}
If this condition is not satisfied, there is no feasible solution under the current fixed variables.
Thus, we can prune this subproblem.
This condition is a necessary condition for the continuous relaxation of the constraint.
Therefore, even if this condition is satisfied, there is a possibility that a feasible solution does not exist.

Next, consider the inequality constraint $\sum_iA_{ri}x_i \leq \tilde{b}_r$.
We can consider two cases.
The first case is when there is no feasible solution.
This case is achieved when the lower bound exceeds $\tilde{b}$.
We can write it as follows:
\begin{equation}
    \tilde{b}_r < \inf\qty(\sum_iA_{ri}x_i).
\end{equation}
In this case, the subproblem can be pruned.
Another case is that the constraint is always satisfied no matter what choice is made in the remaining variables.
This case is when the upper bound is less than $\tilde{b}$.
It can be written as follows:
\begin{equation}
    \sup\qty(\sum_iA_{ri}x_i) \leq \tilde{b}_r.
\end{equation}
This constraint can be ignored in this and subsequent subproblems.

\printbibliography[title=References]

\end{document}